\documentclass{aa}
\usepackage{natbib}
\usepackage{graphicx}
\bibpunct{(}{)}{;}{a}{}{,}
\defcitealias{Moitinho2001a}{Paper~I}
\defcitealias{RNGC}{RNGC}
\defcitealias{Dias2002cat}{DAML02}
\begin{document}
\title{Open clusters in the Third Galactic Quadrant. II.\\
  The intermediate age open clusters NGC 2425 and NGC 2635
  \thanks{Based on observations collected at CTIO and ESO}
  \thanks{Tables~2 and~3 are only available in electronic form at the CDS via
    anonymous ftp to {\tt cdsarc.u-strasbg.fr (130.79.128.5)} or via
    {\tt http://cdsweb.u-strasbg.fr/cgi-bin/qcat?J/A+A//} } }
\author{A. Moitinho\inst{1} \and G. Carraro\inst{2,3,4} \and G.
  Baume\inst{5} \and R.A. V\'azquez\inst{5} }

   \offprints{A. Moitinho\\ \email{andre@oal.ul.pt}}

    \institute{CAAUL, Observat\'orio Astron\'omico de Lisboa, Tapada da Ajuda,
               1349-018 Lisboa, Portugal
           \and
               Departamento de Astron\'omia, Universidad de Chile, Casilla 36-D,
               Santiago, Chile
            \and
               Yale University, Department of Astronomy, PO Box 208101,
               New Haven, CT 06520-8101
            \and
              Dipartimento di Astronomia, Universit\`a di Padova, Vicolo
              Osservatorio 5, I-35122, Padova, Italy
            \and
               Facultad de Ciencias Astron\'omicas y Geof\'{\i}sicas de la
               UNLP, IALP-CONICET, Paseo del Bosque s/n, La Plata, Argentina
               }

   \date{Received **; accepted **}
   
   \abstract{We analyse CCD broad band ($U\!BV(RI)_C$) photometric
     data obtained in the fields of the poorly studied open clusters
     \object{NGC~2425} and \object{NGC~2635}.  Both clusters are found
     to be of intermediate age thus increasing the population of open
     clusters known to be of the age of, or older than, the Hyades.
     More explicitly, we find that \object{NGC~2425} is a 2.2 Gyr old
     cluster, probably of solar metallicity, located at 3.5 kpc from
     the Sun.  \object{NGC~2635} is a Hyades age (600 Myr) cluster
     located at a distance of 4.0 kpc from the Sun.  Its Colour
     Magnitude Diagram reveals that it is extremely metal poor for its
     age and position, thus making it a very interesting object in the
     context of Galactic Disk chemical evolution models.
      }

\authorrunning{Moitinho et al.}
\titlerunning{\object{NGC~2425} and \object{NGC~2635}}

\maketitle
%
%________________________________________________________________

\section{Introduction\label{sec:intro}}

This paper is part of a series motivated by the need for a homogeneous
set of open cluster distances and ages to investigate  the structure
and development of the Milky Way's disk in the
third Galactic quadrant. We are interested in
the interplay between spatial structure and the star
formation history in the Canis Major-Puppis-Vela region of the Galaxy.
The other primary goals are to recover possible extensions of the
Perseus and Norma-Cygnus arms toward the third Galactic quadrant
and to understand the chemical evolution of this poorly-known region.

In \citet[ hereafter \citetalias{Moitinho2001a}]{Moitinho2001a} we
presented a photometric database obtained from a CCD $U\!BV\!RI$
survey of 30 open clusters in the Galactic longitude range
$217^{\circ}< l <260^{\circ}$.  The data presented in
\citetalias{Moitinho2001a} were obtained using a single instrumental
setup (telescope, CCD and filters) and were reduced in the same way,
resulting in a homogeneous photometric database of approximately
65\,000 stars.

Previous papers by our group \citep[][ and references
therein]{Baume2004,Giorgi2005a,Carraro2005b} reported results on
other clusters in this region, namely \object{NGC~2588},
\object{NGC~2580}, \object{NGC~2571}, \object{Pismis~8},
\object{Pismis~13}, \object{Pismis~15}, \object{Ruprecht~4} and
\object{Ruprecht~7}.

We now present the first detailed study of the open clusters
\object{NGC~2425} (=C0736-147) and \object{NGC~2635} (=C0836-345
=Melotte~89 =Collinder~190 =ESO371-01).  Photometric diagrams are
analysed and reddenings, distances and ages are derived, as well as
photometric estimates of the clusters' metallicities.  To the best of
our knowledge, only \object{NGC~2635} has been previously studied by
\citet{Vogt1972} who obtained UBV photoelectric photometry for 6
stars. However, their analysis led them to suggest that there is no
cluster in the direction of \object{NGC~2635}.

In Sect.~\ref{sec:data}, we introduce the observational material used
in this study. Sect.~\ref{sec:counts} is devoted to refining the
present estimates of the clusters' central coordinates and radii. The
photometric analysis is shown in Sect.~\ref{sec:pars}, where the
methods used in the determination of the clusters' fundamental
parameters are also described.  Finally, Sect.~\ref{sec:conc}
summarises our main results and adds some concluding remarks.

\begin{table}
\caption{Central coordinates of the observed objects.\label{tab:coords}}
\fontsize{8} {10pt}\selectfont
\begin{center}
\begin{tabular}{ccccc}
\hline
\hline
\multicolumn{1}{c} {$Name$} &
\multicolumn{1}{c} {$\alpha_{2000}$}  &
\multicolumn{1}{c} {$\delta_{2000}$}  &
\multicolumn{1}{c} {$l$} &
\multicolumn{1}{c} {$b$} \\
\hline
\object{NGC~2425} & 07:38:19.9 & -14:53:15.4 & $231.52^{\circ}$ & $+3.31^{\circ}$ \\
\object{NGC~2635} & 08:38:27.9 & -34:46:40.6 & $255.61^{\circ}$ & $+3.96^{\circ}$ \\
\hline
\hline
\end{tabular}
\end{center}
\end{table}

\section{Data set\label{sec:data}}
\subsection{Optical data \label{sec:opdata}}
\begin{figure}
\begin{center}
\resizebox{0.84\hsize}{!}{\includegraphics{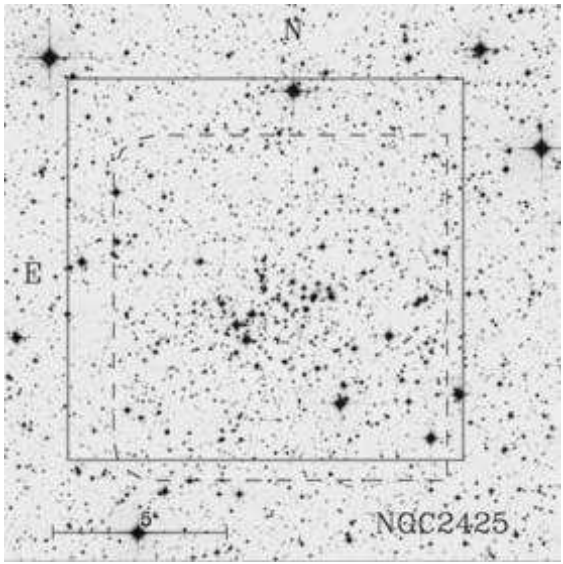}}
\caption{Second generation Digitized Sky Survey (DSS-2), red filter
image of the field of \object{NGC~2425}. The areas covered by the CTIO and ESO
observations are indicated by solid and dashed lines, respectively.
\label{fig:field2425}}
\end{center}
\end{figure}

\begin{figure}
\begin{center}
\resizebox{0.84\hsize}{!}{\includegraphics{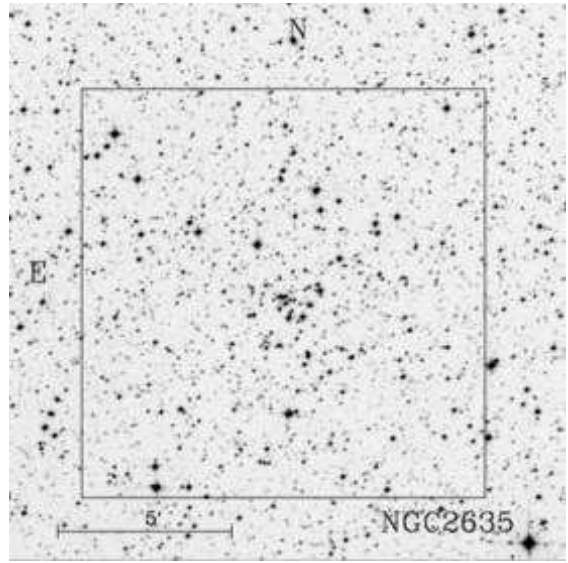}}
\caption{Second generation Digitized Sky Survey (DSS-2), red filter
image of the field of \object{NGC~2635}. The area covered by the CTIO 
observations is indicated by the solid square.\label{fig:field2635}}
\end{center}
\end{figure}

CCD $U\!BV\!RI$ images of \object{NGC~2425} and \object{NGC~2635} were
acquired with the CTIO 0.9m telescope during a run in January 1998.
Observations, reductions, error analysis and comparison with other
photometries are thoroughly described in \citetalias{Moitinho2001a}.
Further CCD $VI$ images were obtained for
\object{NGC~2425} at ESO - La Silla - with the EMMI camera mounted on
NTT on the night of December 9, 2002. The typical seeing was about
$1''$.  The camera has a mosaic of two $2048 \times 4096$ pixels CCDs
which samples a $9\farcm9 \times 9\farcm1$ field. The images were
binned $2 \times 2$, which resulted in a plate scale of
$0\farcs332$/pix. Details on the reductions of these data
are given in \citet{Baume2004}.  The fields covered by CTIO and ESO
observations are shown in Figs.~\ref{fig:field2425}
and~\ref{fig:field2635}.

Combining both data sets required a detailed comparison of their plate
scales and photometric scales. While matching the X,Y pixel coordinates
from both sets, we noticed that no one-to-one correspondence could be
satisfactorily achieved through a simple linear transformation. Indeed,
the residuals of the transformation displayed a jump in the middle of the
NTT X axis. This is likely due to the junction between the two CCDs that
compose the EMMI camera. Taking this effect into account, the NTT
coordinates were transformed without trouble to the system of the CTIO
X,Y positions and sources from both sets were cross-identified. The rms
residuals of the transformations were around 0.15 CTIO pixels (1 CTIO pixel
$\approx 0.39\arcsec$). 

The NTT $VI$ measurements were then linked to the CTIO photometry
through linear transformations of the form $V_{CTIO} - V_{NTT} =
\alpha_0+ \alpha_1 (V-I)_{NTT}$; $(V-I)_{CTIO} =\beta_0 + \beta_1
(V-I)_{NTT}$, using stars with estimated errors of less than 0.015 mag,
which corresponds to $V_{CTIO} < 18$mag. The rms of the
photometric transformations were $\approx 0.02$ mag. Both photometries
were then combined by averaging the measurements weighted by their
errors.  The errors in the CTIO magnitudes are described in
\citetalias{Moitinho2001a}, and are taken to be the dispersion of the
measurements when repeated observations were available, or the errors
output by ALLSTAR in the case of single measurements. For the NTT
magnitudes the adopted errors were those given by ALLSTAR.

\subsection{Infrared data and astrometry\label{sec:irdata}}
Equatorial coordinates were computed for each star by applying linear
transformations, determined for each cluster field, to the CCD X,Y
positions. The transformations were determined in two steps: First,
matched lists of X, Y and RA, DEC were built by visually identifying
about 20-30 TYCHO-2 \citep{Hog2000} and 2MASS stars in each cluster
field. These lists were used in obtaining transformations that yielded
preliminary equatorial coordinates for all measured stars. The second
step was to cross-identify all our sources in common with the same
catalogues by matching the preliminary coordinates to the catalogued
ones using a computer program. This resulted in extended lists of
stars which were used in computing the final transformations. The RMS
of the residuals were $\sim 0.15''$, which is about the astrometric
precision of the 2MASS catalogue, $\sim 0.12''$ \citep{2MASS}, as
expected since most of the coordinates were retrieved from this
catalogue.

Cross-identification of our stars with 2MASS point sources allowed us
to build a catalogue of $U\!BV\!RIJHK$ photometry which constitutes
the main observational base used in this study. The photometric data
for \object{NGC~2425} and \object{NGC~2635} are given in tables~2
and~3, respectively\addtocounter{table}{2}. These tables include X,Y
CCD coordinates; $\alpha_{2000}$ and $\delta_{2000}$ equatorial
coordinates, $U\!BV\!RI$ magnitudes and their estimated errors.
tables~2 and~3 are only available electronically either at the CDS or
at the WEBDA\footnote{http://obswww.unige.ch/webda/navigation.html}
sites.

\section{Star counts, cluster centres and sizes\label{sec:counts}}
As a first step, we estimated the position of each cluster's centre
following a variation of the method given in \citet{Moitinho1997}: the
surface stellar distributions of stars brighter than $V = 19$ were
convolved with a $100 \times 100$ pixel ($\approx 0.66^{\prime}
\times 0.66^{\prime}$) Gaussian kernel to smooth out the details of
the spatial distributions.  The derived cluster centres, taken to be
the geometrical centres of the regions of enhanced stellar density,
are given in table~\ref{tab:coords}.  Both values lie very close to the
ones given by \citet{Dias2002cat}.

\begin{figure}
\resizebox{\hsize}{!}{\includegraphics{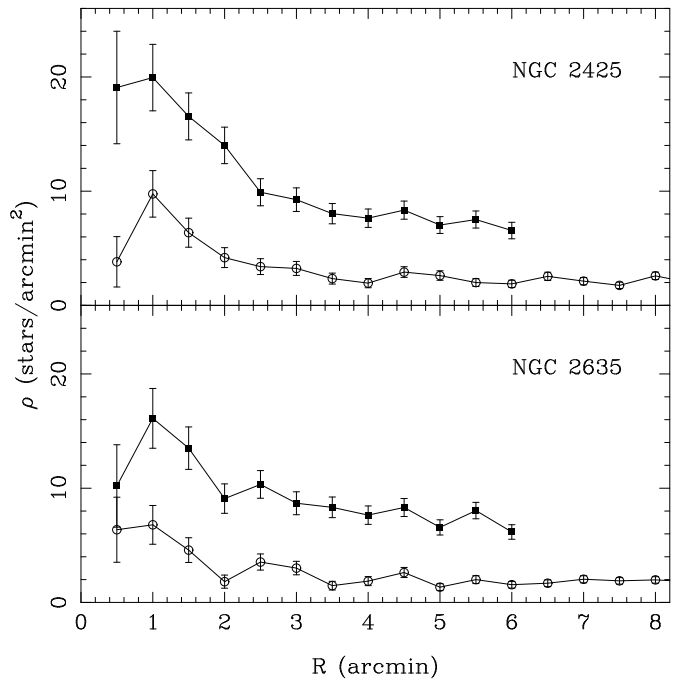}}
\caption{Radial density profiles for \object{NGC~2425} (upper panel) and 
  \object{NGC~2635} (lower panel). Filled squares: CCD data.  Open circles:
  2MASS data. Poisson error bars ($1\sigma$) are also shown.
\label{fig:radial}}
\end{figure}

Once a cluster's centre was determined, its radial density profile was
built by counting stars in a number of successive rings, $0.5'$ wide,
and dividing the counts by the correspondent ring's area.  Appropriate
corrections were applied to the areas of rings not totally included in
the image. Density profiles were built for both the optical data (for
$V < 19$ mag) and for the 2MASS infrared data (for $K < 14.5$ mag).
The magnitude cut-offs were roughly selected to reduce
contamination by field stars and possible artifacts due to
incompleteness (especially true for the 2MASS data).  The radial
density profiles are shown in Fig.~\ref{fig:radial}.  In the case of
\object{NGC~2425} (upper panel), we notice an obvious stellar
density enhancement that appears to reach the background level at
$\sim 4^{\prime}$. For \object{NGC~2635} (lower panel), the probable
radius is also found to be $\sim 4.0^{\prime}$.  These star counts
yield radii that are significantly larger than the visual estimates
reported in \citet{Dias2002cat}.  In both cases, the radii are taken
to be the range beyond which the star counts follow a flat
distribution, describing the field star population.

\begin{figure}
\centering
\includegraphics[width=8.8cm]{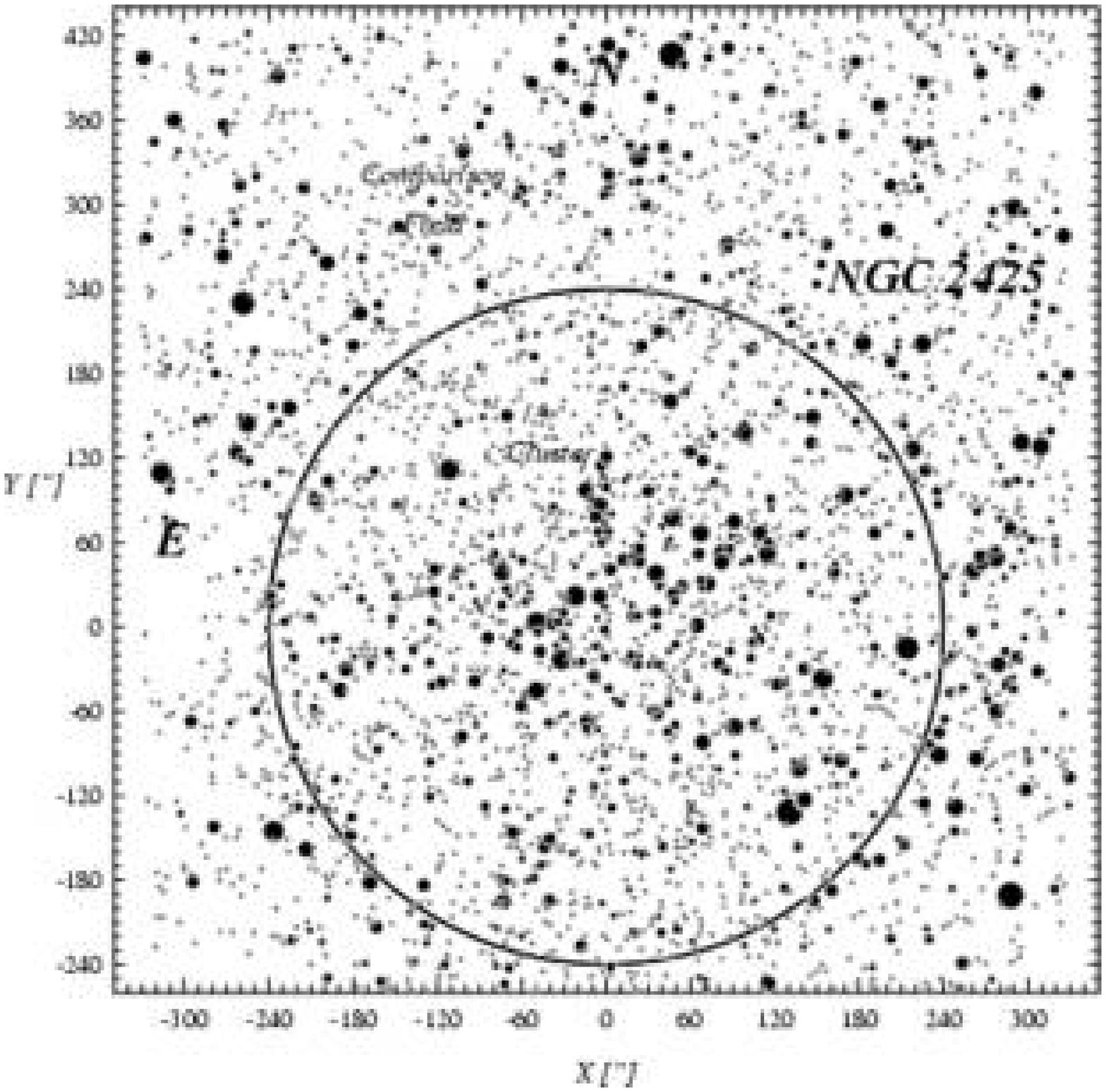}
\caption{Finding chart for the observed region around \object{NGC~2425}
  ($V$ filter). The solid circle, $4\farcm0$ indicates the derived
  cluster limits (see Sect.~\ref{sec:counts} and
  Fig.~\ref{fig:radial}). The adopted comparison field is the region
  beyond the cluster limits.  The origin ($X=0$; $Y=0$) corresponds to
  the cluster central coordinates ($\alpha_{2000} =$ 07:38:19.9;
  $\delta_{2000} =$ -14:53:15.4).  $X,Y$ are given in
  arcsecond.\label{fig:find2425}}
\end{figure}

\begin{figure}
\centering
\includegraphics[width=8.8cm]{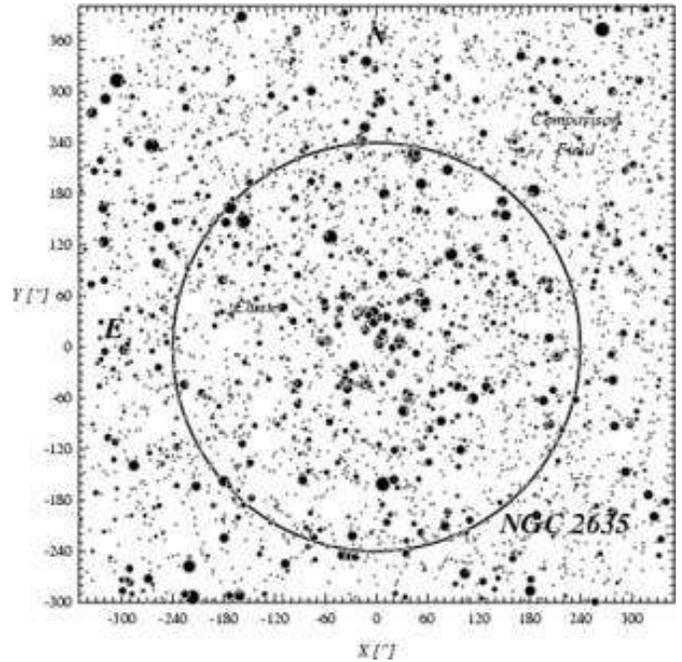}
\caption{Finding chart for the observed region around \object{NGC~2635}
  ($V$ filter). The solid circle, $4\farcm0$ indicates the derived
  cluster limits (see Sect.~\ref{sec:counts} and
  Fig.~\ref{fig:radial}). The adopted comparison field is the region
  beyond the cluster limits.  The origin ($X=0$; $Y=0$) corresponds to
  the cluster central coordinates ($\alpha_{2000} =$ 08:38:27.9;
  $\delta_{2000} =$ -34:46:40.6).  $X,Y$ are given in
  arcsecond.\label{fig:find2635}}
\end{figure}

The derived centres and radii are again shown in the finding charts in
Figs.~\ref{fig:find2425} and~\ref{fig:find2635} for \object{NGC~2425} and
\object{NGC~2635}, respectively.  These plots also indicate the areas adopted
as control fields for the analysis of the photometric diagrams
performed in the next sections.

\section{Photometric analysis\label{sec:pars}}
\begin{figure}
\centering
\includegraphics[width=8.8cm]{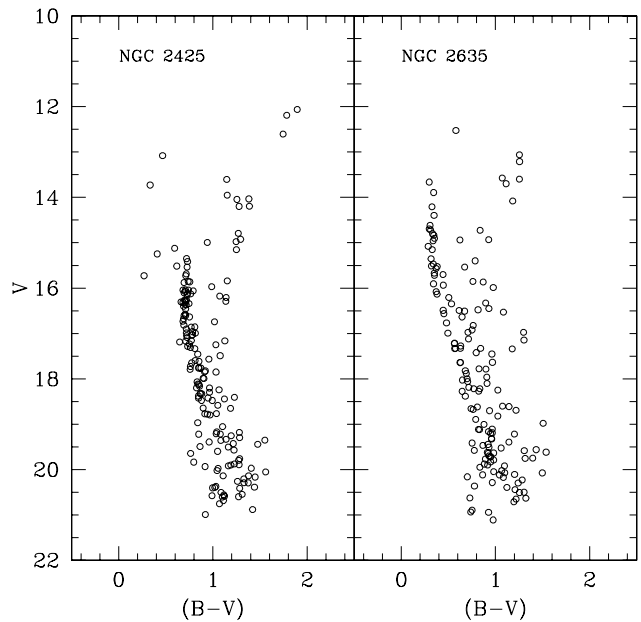}
\caption{CMDs of \object{NGC~2425} (left
  panel) and of \object{NGC~2635} (right panel). Both diagrams display
  only the stars within the central $R < 1.5'$. See details in
  Sect.~\ref{sec:pars}.\label{fig:cmd0}}
\end{figure}

To estimate the cluster ages we start by considering the
CMDs in the V vs (B-V) plane shown in Fig.~\ref{fig:cmd0}.  Only stars
located within the very inner central regions were plotted to
reduce the effect of field star contamination and better interpret
the shape of the cluster sequences while still containing a reasonable
number of cluster stars. The circular extractions plotted in
Fig.~\ref{fig:cmd0} are $R < 1.5'$ for both \object{NGC~2425} and
\object{NGC~2635}.

The CMD of \object{NGC~2425} is shown in the left panel of
Fig.~\ref{fig:cmd0}.  It is an old open cluster with a turn-off (TO)
at $V\approx$16 and a clump of red stars at $V\approx$14.1. A rough
age estimate, using the calibration of \citet{Carraro1994a} for
$\Delta V$ (the magnitude difference between the TO and red clump),
yields 2.5 Gyr. As discussed by those authors, the estimate is
independent of the distance modulus and reddening.  Like other
clusters of about this age (e.g. \object{NGC~2420}, \object{NGC~2204})
the red giant branch (RGB) is barely visible. The Main Sequence (MS)
extends for 5 mag up to $V\approx 20$, and gets wider at fainter
magnitudes, which we ascribe to increasing photometric errors at
increasing magnitude, and to contamination by field stars.

The CMD of \object{NGC~2635} shown in the right panel of
Fig.~\ref{fig:cmd0} reveals a younger cluster, with a TO at about
$V\approx 15$, and a wide clump, typical of moderate age clusters.  The
fact that, except for the clump stars, the cluster sequence is fainter
than $V\approx 14$ explains why \citet{Vogt1972} did not find signs of a
cluster in their photoelectric study.  Using again the
\citet{Carraro1994a} calibration, we infer an age lower than 1 Gyr.
Due to the younger age, the MS is in this case slightly more extended in
magnitude. A parallel sequence, probably due to binary stars, is also
visible redward of the MS.

The following sections are devoted to the first determination of
these clusters' fundamental parameters (reddening, distance and age)
through a detailed analysis of their photometric diagrams.

\subsection{\object{NGC~2425}\label{2425}}

\begin{figure*}
\centering
\includegraphics[height=8.8cm]{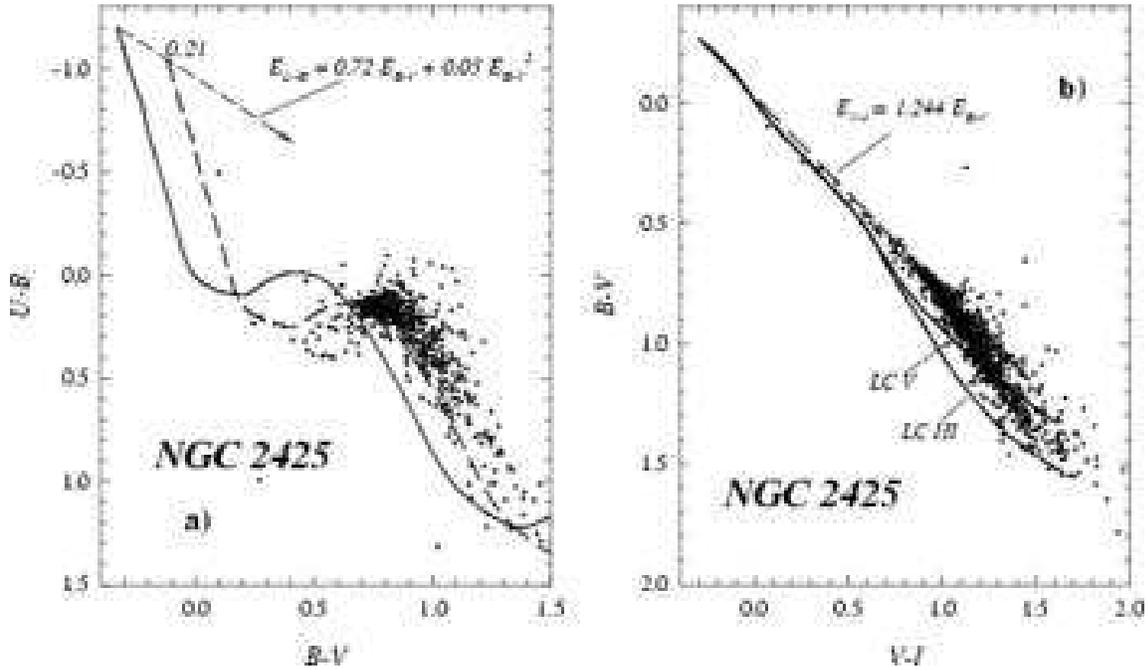}
\caption{Optical colour-colour diagrams for the stars located within the 
  radius of \object{NGC~2425} {\bf a)} $U-B$ vs. $B-V$ diagram.  The
  solid line is the \citet{Schmidt-Kaler1982} ZAMS, whereas the dashed
  line is the same ZAMS, but shifted by the derived colour excess. The
  dashed arrow indicates the normal reddening path. {\bf b)} $B-V$ vs.
  $V-I$ diagram.  Lines have the same meaning as in panel a) except
  that the reference lines are from \citet{Cousins1978a}.
\label{fig:ccd2425}}
\end{figure*}

\begin{figure*}
\centering
\includegraphics[height=18cm]{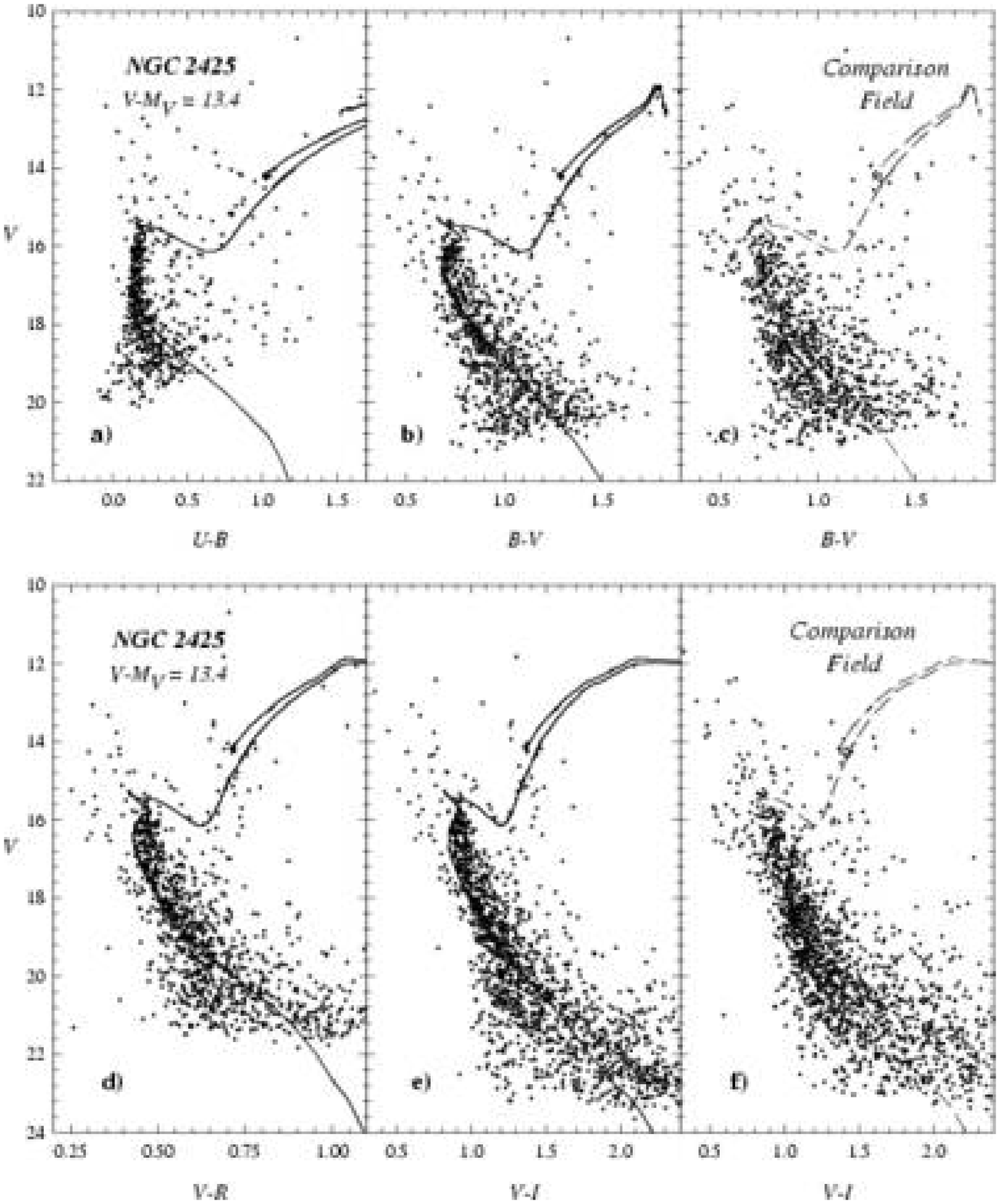}
\caption{Optical colour-magnitude diagrams for the stars located inside the
  radius of \object{NGC~2425} and in the adopted comparison field. The
  solid lines are the \citet{Schmidt-Kaler1982} and
  \citet{Cousins1978a} empirical ZAMS (for $U-B$ and $B-V$) and MS
  (for $V-R$ and $V-I$) respectively, and the dashed lines are the
  \citet{Girardi2000iso} 2.2 Gyr solar metallicity isochrone, all
  compensated for the effects of reddening and distance. The derived
  apparent distance modulus is $V-M_{V} = 13.4$
  ($V_{0}-M_{V}=V-M_{V}-3.1E_{B-V}=12.75$).
\label{fig:cmd2425}}
\end{figure*}

The Two Colour Diagrams (TCDs) and Colour Magnitude diagrams (CMDs) of
\object{NGC~2425} are presented in Figs.~\ref{fig:ccd2425}
and~\ref{fig:cmd2425}, respectively.  In Fig.~\ref{fig:cmd2425},
panels c) and d) represent the {\it Comparison Field} (CF) CMDs (see
Fig.~\ref{fig:find2425}). This field was chosen to
cover the same area as the cluster and to be within the region
covered by both CTIO and NTT observations.

Because of its old age, the stars in \object{NGC~2425} have too late
spectral types, which make the TCDs not very useful in deriving
memberships. We can, however, use those diagrams to infer the cluster
reddening and reddening path.  A reddening of $E(B-V)=0.21$ is derived
by fitting the Schmidt-Kaler (1982) ZAMS to the cluster sequence in
the $U-B$ vs $B-V$ plane, and the Cousins (1978) reference line to the
sequence in the $V-I$ vs $B-V$ plane. The TCDs show that the reddening
path is the normal one, as also previously indicated in
\citetalias{Moitinho2001a} for this region of the Galaxy.

CMDs for several different colour combinations are shown in
Fig.~\ref{fig:cmd2425}.  The different CMDs do not reach the same
limiting magnitude, $V_{lim}$, ranging from $\sim 19$ in the diagrams
involving $U-B$ to $\sim 23$ in the $V$ vs $V-I$ plane. This is due to
the different wavelength sensitivity of the detector/filter
combinations and to the the deeper $V,I$ NTT data (see
Sect.~\ref{sec:opdata}).  Superimposed on each diagram is the same 2.2
Gyr solar composition isochrone from the Padova group
\citep{Girardi2000iso}. The isochrone nicely reproduces the observed
cluster sequence (MS, TO, RGB) in all the diagrams.  Other
combinations of age and metallicity have been tried, but none provided
such a good fit.

The isochrone fit yields an apparent distance modulus $(m-M)=13.4$.
This value together with $E(B-V)=0.21$, as derived from the TCDs, puts
\object{NGC~2425} at 3550 pc from the Sun, which corresponds to a
Galactocentric distance of 10.7 kpc, 200 pc above the Galactic plane.
 
\begin{figure*}
\centering
\includegraphics[height=9cm]{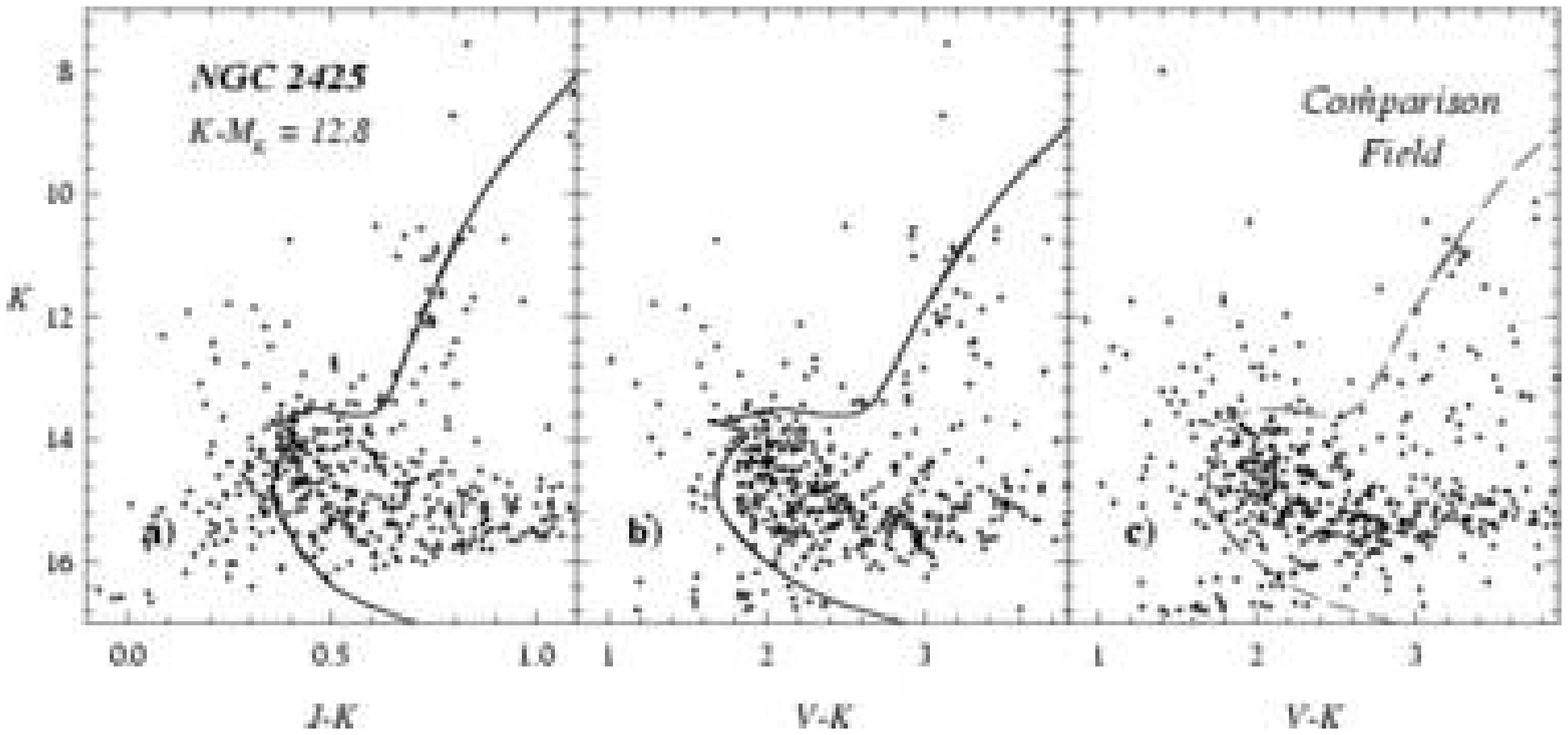}
\caption{Composite 2MASS/Optical colour-magnitude diagrams of stars 
  located inside the radius of \object{NGC~2425} and in the adopted
  comparison field.  Symbols and lines have the same meanings as in
  Fig.~\ref{fig:ccd2425}.\label{fig:cmd2mass2425}}
\end{figure*}

Fig.~\ref{fig:cmd2mass2425} shows the best fit isochrone plotted over
the infrared 2MASS CMDs. The combination of age and distance results in
a cluster sequence that is too faint to be clearly identified in these
diagrams. As will be shown in Sect.~\ref{sec:2635}, that is not the
case of the infrared diagrams of \object{NGC~2635}.

\subsection{\object{NGC~2635}\label{sec:2635}}

\begin{figure*}
\centering
\includegraphics[height=9cm]{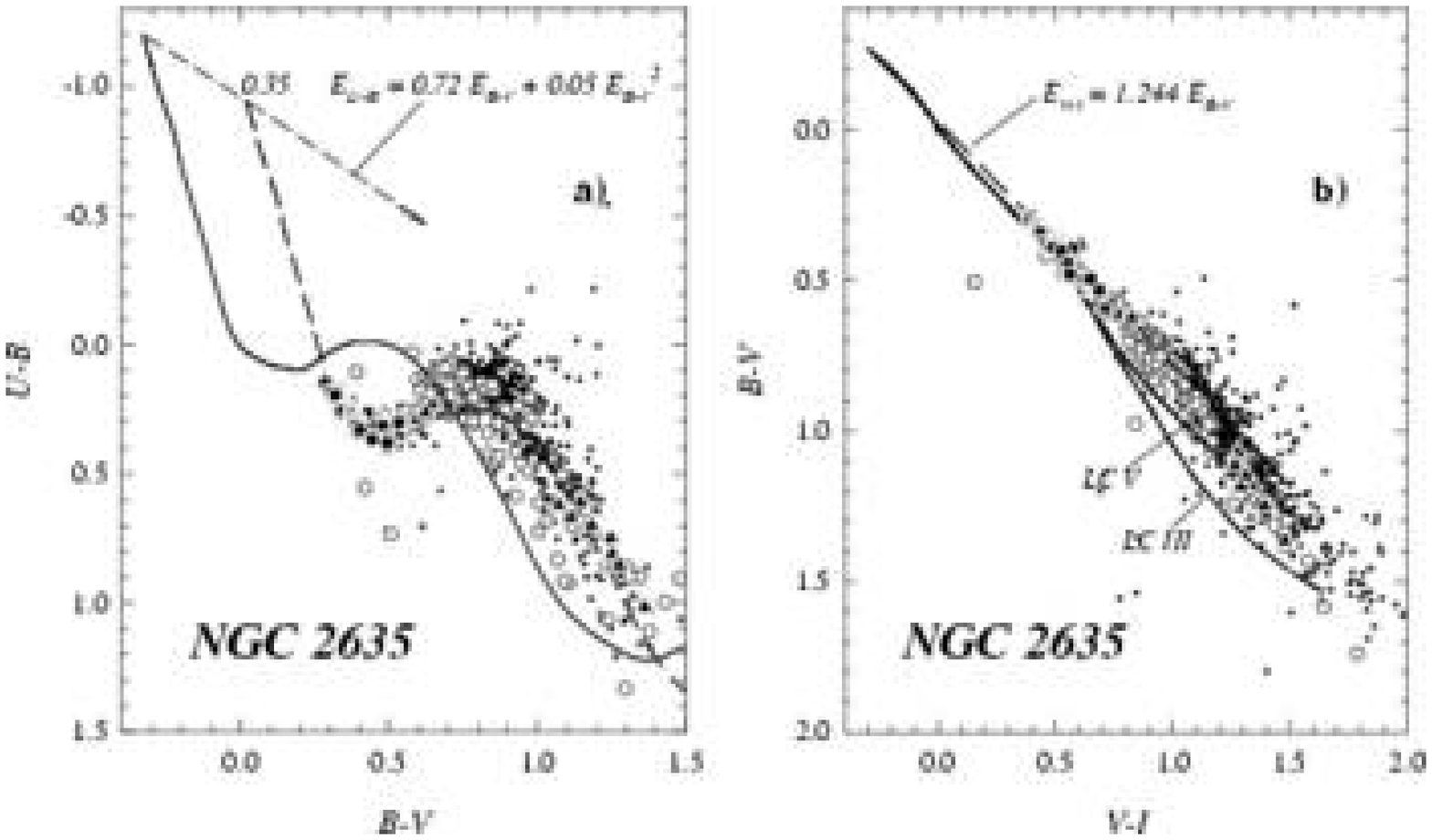}
\caption{Optical colour-colour diagrams for the stars located within 
  the radius of \object{NGC~2635} {\bf a)} $U-B$ vs. $B-V$
  diagram. Black circles are
  likely member stars ($lm$), black triangles are probable member
  stars ($pm$), white circles are non-member stars ($nm$), and dots
  are stars without any membership assignment. The solid line is the
  \citet{Schmidt-Kaler1982} ZAMS,  the dashed one is the same
  ZAMS, but shifted by the derived colour excess. The dashed arrow
  indicates the normal reddening path. {\bf b)} $B-V$ vs. $V-I$
  diagram.  Symbols and lines have the same meaning as in panel a)
  except that the reference lines are from \citet{Cousins1978a}}
\label{fig:ccd2635}
\end{figure*}

\begin{figure*}
\centering
\includegraphics[height=18cm]{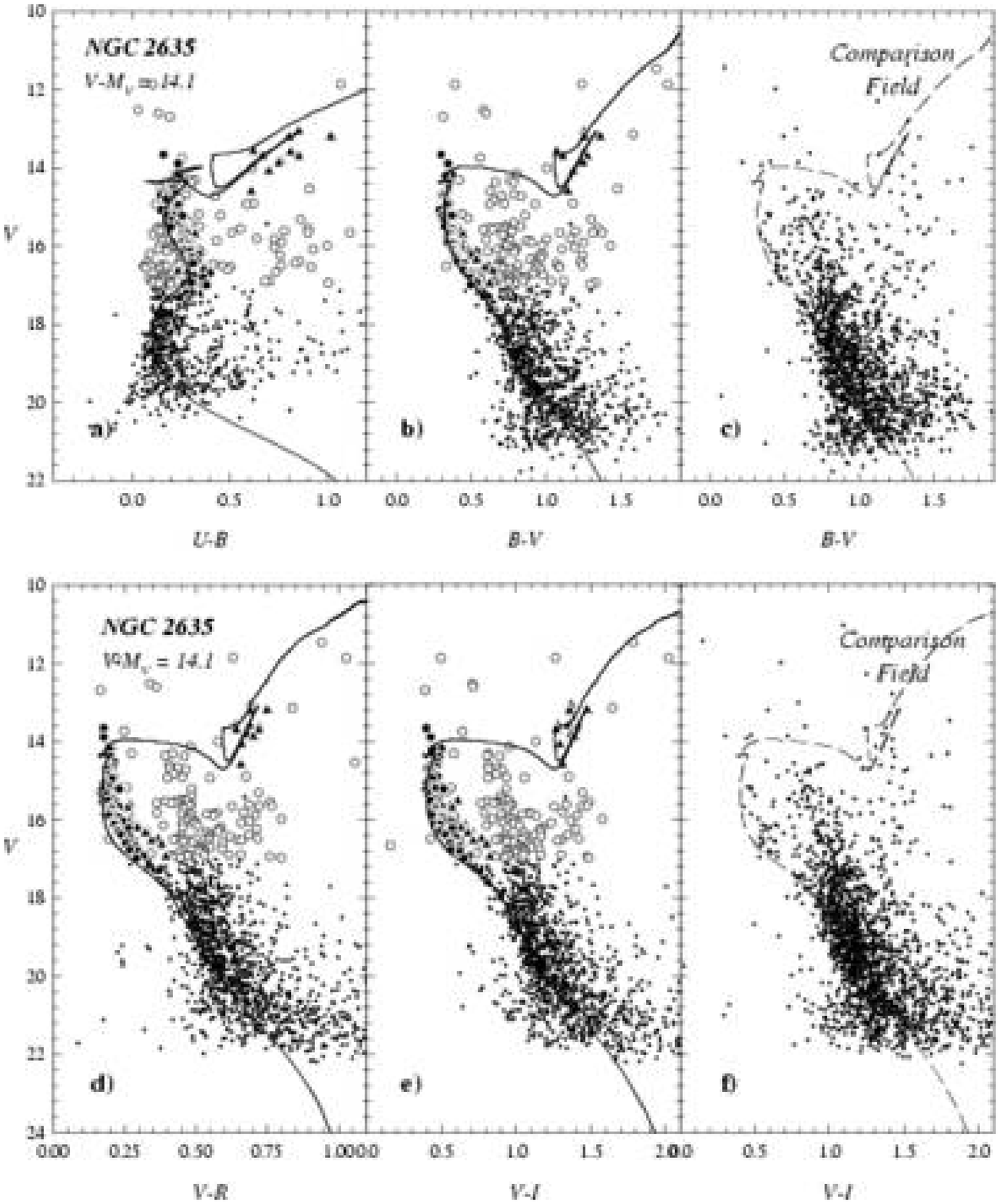}
\caption{Optical colour-magnitude diagrams for the stars located within
  the adopted radius of \object{NGC~2635} and in the adopted
  comparison field.  Symbols have the same meaning as in
  Fig.~\ref{fig:ccd2635}. The dashed and dotted lines are the
  \citet{Girardi2000iso} 600 million year isochrone, calculated for
  Z=0.004, and compensated for the effects of reddening and distance.
  The derived apparent distance modulus is $V-M_{V} = 14.1$
  ($V_{0}-M_{V}=V-M_{V}-3.1E_{B-V}=13.02$).
\label{fig:cmd2635}}
\end{figure*}

Unlike \object{NGC~2425}, the presence of earlier type stars in
\object{NGC~2635} allows us to perform a photometric membership search
following the technique described in \citet{Baume2004}.  The
method is based on a close synoptic scrutiny of the locations of the
stars in TCDs and CMDs built using several colour combinations. The
results are illustrated in Figs.~\ref{fig:ccd2635}
and~\ref{fig:cmd2635}.  These diagrams display a clear cluster
sequence above $V \approx 17.5-18.0$. So the individual stellar
locations down to this magnitude are examined in all the photometric
diagrams in a fashion similar to that in \citet{Baume2004}:
\begin{itemize}
\item For stars brighter than $V \sim 16.0$, if they have coherent
  locations near the ZAMS, they are considered likely members ($lm$).
\item Fainter stars, with magnitudes in the range $V \sim 16.0-18.0$ in
  the same conditions are considered only probable members ($pm$).
\item If some stars are brighter than $V \sim 16.0$, well placed in
  the TCDs of Fig.~\ref{fig:ccd2635} but located slightly over the ZAMS
  in Fig.~\ref{fig:cmd2635}; they are still considered as $pm$ since
  their brighter magnitude  could be due to binarity.
\item The number of stars in each magnitude bin is checked for
  agreement with the counts obtained from the difference
  between the $'cluster~fields'$ and the corresponding $CF$.
\end{itemize}
At fainter magnitudes, contamination from the Galactic field
population (compare with the $CF$ CMDs) becomes severe, preventing
a firm identification of cluster members.

\begin{figure*}
\centering
\includegraphics[height=9cm]{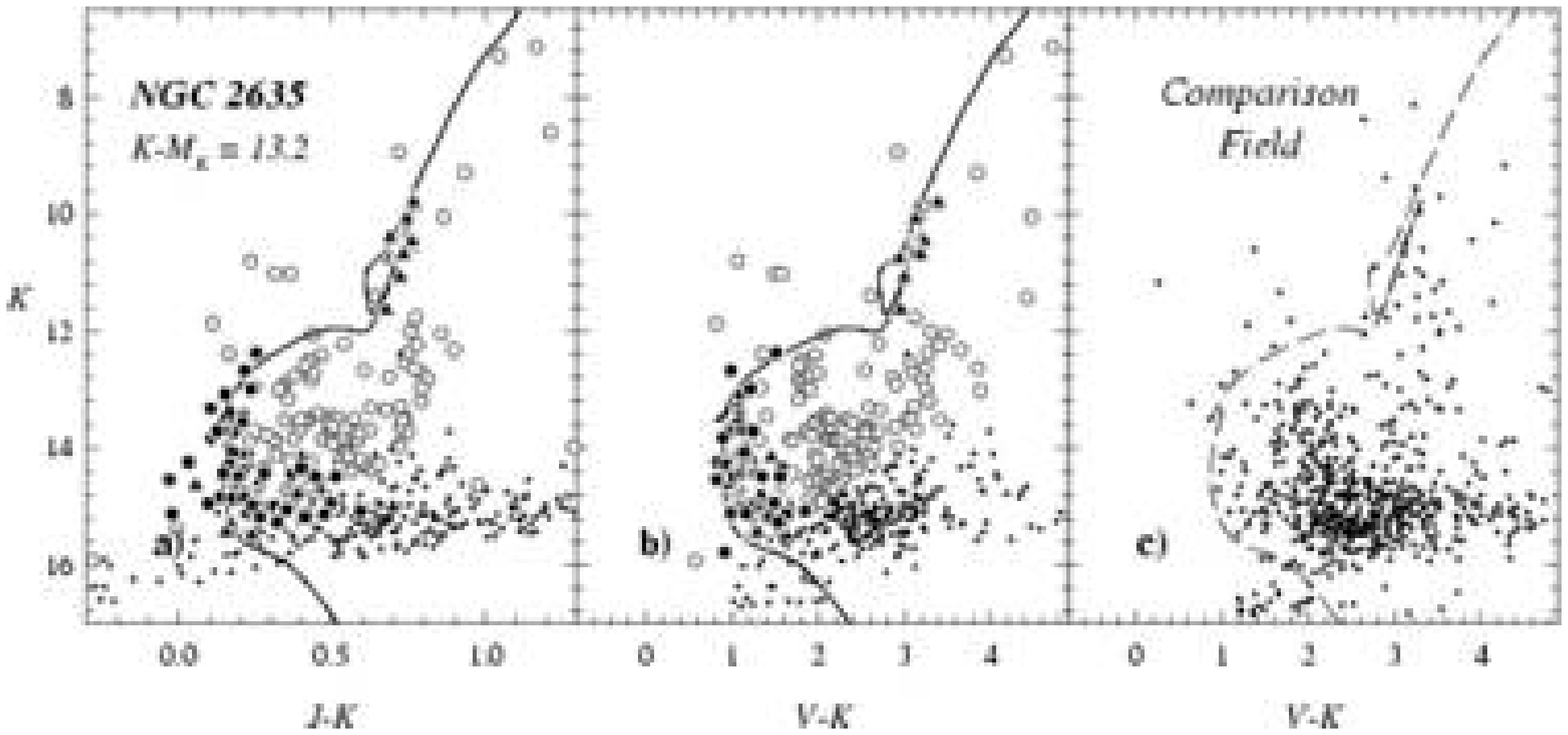}
\caption{Composite 2MASS/Optical colour-magnitude diagrams of stars 
  located inside the radius of \object{NGC~2635} and in the adopted
  comparison field.  Symbols and lines have the same meanings as in
  Fig.~\ref{fig:ccd2635}.\label{fig:cmd2mass2635}}
\end{figure*}

Overplotted on the CMDs of Fig.~\ref{fig:cmd2635} is a 600 million
year old isochrone computed for Z=0.004 \citep{Girardi2000iso},
shifted for a colour excess of $E(B-V)=0.35$ and an apparent distance
modulus of $(m-M)=14.1$.  We have found these parameters to provide
the best fit to all the CMDs simultaneously.  The derived reddening
and apparent distance modulus situate \object{NGC~2635} at a
heliocentric distance of approximately 4 kpc, which corresponds to a
Galactocentric distance of 9.5 kpc, 300 pc above the Galactic plane.
The TCD positions of the earlier spectral type stars (see
Fig.~\ref{fig:ccd2635}) confirm the derived age.  We do, however,
notice the presence of two bright stars above the TO, which are
members according to our photometric criteria, but which are not
fitted by the isochrone.  These two photometric members lie well
within the cluster radius, so they could be blue stragglers or simple
binaries.  The isochrone fit is also shown in the infrared diagrams of
Fig.~\ref{fig:cmd2mass2635}.  Although the $K$ vs $J-K$ CMD presents a
large scatter, mainly due to the precision limit of the 2MASS data,
the combined optical-infrared $K$ vs $V-K$ diagram shows a tighter
sequence that is well reproduced by the model and derived parameters,
with no counterpart in the CF diagram.  Also, in both cluster diagrams
the bright red clump is very well reproduced.

\begin{figure}
\centering
\includegraphics[height=9cm]{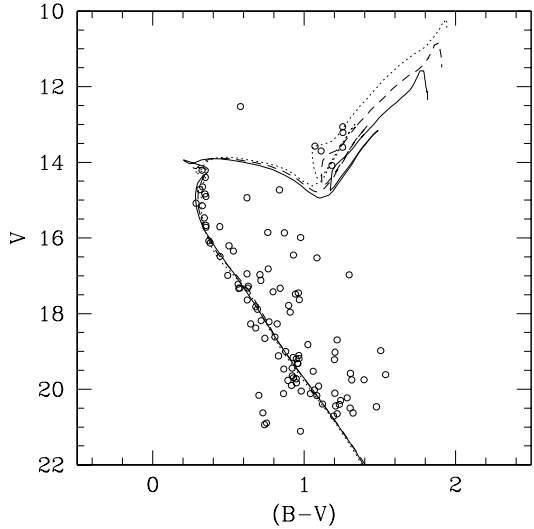}
\caption{Three 600 million year isochrones with different metallicities fitted
  to the sequence of \object{NGC~2635} in the $V$ vs $B-V$ plane, for
  a circular extraction of $1.5'$.  The solid, dashed and dotted lines
  are Z=0.019, Z=0.008 and Z=0.004, respectively.
\label{fig:metal}}
\end{figure}

The need to adopt such a low metallicity is better illustrated in
Fig.~\ref{fig:metal}, where three isochrones for the same age, but
with different metallicities, are plotted.  Although all three
isochrones reproduce well the shape of the TO, only the low
metallicity curve simultaneously fits both the cluster sequence and
the red clump for our derived reddening. Adjusting the
parameters, we find that for a solar metallicity an age of 1 Gyr would
fit the data with the untenable reddening value of 0.0 mag, while the
intermediate metallicity isochrone would work for an age of 0.8 Gyr
and a reddening of 0.15 mag, too low to be compatible with the
findings from the TCDs.

\section{Discussion and conclusion\label{sec:conc}}

We have presented the first CCD photometric study of the open clusters
\object{NGC~2425} and \object{NGC~2635}. The two clusters were found
to be of intermediate age, allowing us to update the census of the old
open cluster population in the Galactic disk. The basic cluster
parameters derived in this study are presented in
Tabs.~\ref{tab:coords} and~\ref{tab:pars}.

\begin{table}
\caption{Derived parameters for \object{NGC~2425} and 
\object{NGC~2635}.\label{tab:pars}}
\fontsize{8} {10pt}\selectfont
\begin{center}
\begin{tabular}{cccccc}
\hline
\hline
\multicolumn{1}{c}{$Name$} &
\multicolumn{1}{c}{$E(B-V)$}  &
\multicolumn{1}{c}{$(V-M_V)$}  &
\multicolumn{1}{c}{$Dist$} &
\multicolumn{1}{c}{$Age$} &
\multicolumn{1}{c}{$radius$} \\
& 
\multicolumn{1}{c}{(mag)} & 
\multicolumn{1}{c}{(mag)} & 
\multicolumn{1}{c}{(Kpc)}&
\multicolumn{1}{c}{(Myr)} &
\multicolumn{1}{c}{(')} \\
\hline
\object{NGC~2425} & 0.21 & 13.4 & 3.55 & 2200 & 4.0 \\
\object{NGC~2635} & 0.35 & 14.1 & 4.00 &  600 & 4.0 \\
\hline
\hline
\end{tabular}
\end{center}
\end{table}

We found that \object{NGC~2425} is a 2.2 Gyr old
cluster, located at a Galactocentric distance of 10.7 kpc.  The
overall good isochrone fit through the stellar sequences in several
CMDs, using various colour combinations, allowed us to conclude that
the cluster has a nearly solar metal abundance.

\object{NGC~2635} is a much younger cluster, with an age of 600 Myr,
close to the age of the Hyades.  We situate this cluster at a
Galactocentric distance of 9.5 kpc, and propose that it has to be
rather metal poor. The metallicity which provides the best isochrone
fit is Z=0.004, which translates into [Fe/H]=-0.61.  Clearly, both
clusters are very interesting objects in the context of the Galactic
disk's chemical evolution, in particular the role they can play in
shaping the radial abundance gradient and the age-metallicity relation
in the Galactic disk \citep{Carraro1998}.  Recent suggestions have
been made that the radial abundance gradient flattens with time
\citep[][, Fig.~3]{Friel2002}.  \object{NGC~2425}, with an age of 2.2
Gyr and solar abundance confirms the slope of the gradient
at intermediate ages (2 to 4 Gyrs).

However, \object{NGC~2635} with an age of half a Gyr and a metallicity
of [Fe/H]=-0.61, significantly deviates from the mean trend.  In a
recent work, \citet{Carraro2005a} studied two intermediate-age open
clusters, \object{NGC~6404} and \object{NGC~6583}, located at about
6.5 kpc from the Galactic centre, which have a probable solar
abundance. If we put the two clusters and \object{NGC~2635} in the
radial gradient defined by the youngest clusters, we infer that the
slope is much steeper than that found by \citet{Friel2002}, thus
casting some doubt on the flattening of the gradient.  Clearly, more
young clusters are needed, especially to better define the distance baseline.

The development of this third Galactic quadrant project
should help clarifying this issue, thus providing more robust constraints
for Galactic disk chemical evolution models \citep{Tosi1996}.

\begin{acknowledgements}
  AM acknowledges financial support from FCT (Portugal) through grants
  BPD/20193/99 and SFRH/BPD/19105/2005 and through the YALO project
  (CTIO observations).  GC acknowledges the support of {\it
    Fundaci\'on Andes}.  GB acknowledges the cooperative international
  agreement Argentino-Italiano SECYT-MAE (IT/PA03 - UIII/077). This
  research made use of the NASA Astrophysics Data System, of the
  Simbad database operated at the Centre de Donn\'es Stellaires ---
  Strasbourg, France, and of the WEBDA open cluster database of J.-C.
  Mermilliod. This publication also makes use of data products from
  the Two Micron All Sky Survey, which is a joint project of the
  University of Massachusetts and the Infrared Processing and Analysis
  Center/California Institute of Technology, funded by the National
  Aeronautics and Space Administration and the National Science
  Foundation.
\end{acknowledgements}

\bibliography{newbib,biblio,tese}
\bibliographystyle{apj}

\end{document}